# MULTIPLE WAVELET COHERENCY ANALYSIS AND FORECASTING OF METAL PRICES

EMRE KAHRAMAN[*,‡] and GAZANFER ÜNAL[†]

[*] Yeditepe University Financial Economics
Financial Economics Graduate and PhD Programs, PhD Candidate
İnönü Mahallesi, Kayışdağı Caddesi, 26 Ağustos Yerleşkesi Ataşehir 34755 İstanbul
[‡]emre.kahraman1@std.yeditepe.edu.tr
[†]Yeditepe University Financial Economics
Financial Economics Graduate and PhD Programs, Coordinator
İnönü Mahallesi, Kayışdağı Caddesi, 26 Ağustos Yerleşkesi Ataşehir 34755 İstanbul
gunal@yeditepe.edu.tr

The assessment of co-movement among metals is crucial to better understand the behaviors of the metal prices and the interactions with others that affect the changes in prices. In this study, both Wavelet Analysis and VARMA (Vector Autoregressive Moving Average) models are utilized. First, Multiple Wavelet Coherence (MWC), where Wavelet Analysis is needed, is utilized to determine dynamic correlation time interval and scales. VARMA is then used for forecasting which results in reduced errors.

The daily prices of steel, aluminium, copper and zinc between 10.05.2010 and 29.05.2014 are analyzed via wavelet analysis to highlight the interactions. Results uncover interesting dynamics between mentioned metals in the time-frequency space. VARMA (1,1) model forecasting is carried out considering the daily prices between 14.11.2011 and 16.11.2012 where the interactions are quite high and prediction errors are found quite limited with respect to ARMA(1.1). It is shown that dynamic co-movement detection via four variables wavelet coherency analysis in the determination of VARMA time interval enables to improve forecasting power of ARMA by decreasing forecasting errors.

## 1. Introduction

Wavelet analysis is becoming a common tool for analyzing localized variations of power within a time series. While major part of economic time series analysis is done either in time or frequency domain separately, two fundamental approaches are combined with wavelet analysis which allows the study in the time frequency domain. By decomposing a time series into time frequency space, one is able to determine both the dominant modes of variability and how those modes vary in time. The wavelet transform has been used for numerous studies in geophysics, including tropical convection [1], the El Niño–Southern Oscillation [2-3], atmospheric cold fronts [4], central England temperature [6], the dispersion of ocean waves [7], wave growth and breaking [8], and coherent structures in turbulent flows [9]. A complete description of geophysical applications can be found in Wavelets in Geophysics [10], while a theoretical treatment of wavelet analysis is given by Daubechies [12]. Finally, Aguiar-Conraria and Soares [44] worked on applications in finance recently. Aguiar-Conraria and Soares [44] also introduced the multiple wavelet coherence (MWC) concept and its application for three variables.

The fundamental advantage of wavelet analysis is the fact that the combination of time and frequency domain time series analysis could be performed. In other word, while most of the time researches focused on either time or frequency domain, it is possible to combine both with wavelet analysis.

The study of co-movement of steel, aluminium, copper and zinc prices using wavelet analysis is utilizing various approaches including discrete and continuous wavelet transform, de-noising, phase coherence and multiple wavelet coherence. The trend and noise are analyzed separately in order to detect the long and short term relationships between the time series and analysis are extended up to four variables case. The main reason to extend the study to four variables is to detect all possible interrelations between variable and to improve the forecasting power in that way. Those selected four metals are the main contributors at non-ferrous metals price index and have significant governance on the metal sector. Therefore, they are chosen in this study to be analyzed.

In addition, the application of multiple wavelet coherence (MWC) is demonstrated. While partial wavelet coherence (PWC) is a technique similar to partial correlation that helps identify the resulting wavelet transform coherence (WTC) between two time series after eliminating the influence of their common dependence, MWC is useful in seeking the resulting WTC of multiple independent variables on a dependent one. MWC is used to determine co-movement or contagion during the study

Researchers are motivated to enlarge the model class to the multivariate case with a successful use of univariate ARMA models for forecasting. It is expected to improve the forecasting power and precision via using more information by including more interrelated variables in the model. Granger's [19] influential definition of causality is actually coming from this idea.

But, it is understood that the process of generalizing univariate models to multivariate ones is not an easy in the ARMA case. Quenouille [20] is the first person who considered multivariate VARMA models. However, it was clear that the specification and estimation of such models was much more difficult with respect to ARMA. The success of the Box-Jenkins [21] modelling strategy for univariate ARMA models in the 1970's lead to further studies of using the corresponding multivariate models and developing estimation and specification strategies. The specific investigations focused on the possibility of using autocorrelations, partial autocorrelations and cross correlations between the variables for model specification. Since in the univariate Box-Jenkins approach, modelling strategies based on such quantities had been successful, it was logical to search for multivariate extensions. Tiao & Box [22], Tiao & Tsay [23-24] Tsay [25-26] Wallis [27], Zellner & Palm [28], Granger & Newbold [29] and Jenkins & Alavi [30] are the main examples of such trials.

On the other hand, these strategies were affording hope for very small systems of two or three variables. Moreover, because of the fact that VARMA representations are not unique (identical), the most useful setup of multiple time series models was under discussion. Critical discussions of the related problems are made by Hannan [31, 32, 33, 34], Dunsmuir & Hannan [35] and Akaike [36]. In the late 1980's, Hannan & Diestler [37] introduced general solution to the structure theory for VARMA models. Furthermore, the development of complete specification strategies was contributed via understanding the structural problems. Lütkepohl [38] gave a recent overview of forecasting with VARMA processes.

VARMA utilizes the information not only in the past values of particular variable of interest but also allow for information in other related variables. VARMA model is utilized within the study to see the positive effect of co-movement between steel, aluminium, copper and zinc prices in forecasting process. The first objective is to determine the highest correlation time interval via wavelet analysis to execute further study on this period. The forecasting powers and precision levels of univariate ARMA and VARMA models are compared by MSE analysis to show that vector analysis, where cross correlations are also included, reduce the forecasting errors.

In Section 2, first multiple wavelet and complex coherence concepts are described and then the calculations are extended to four variables which is one of the main contribution of the paper. In Section 3, wavelet data analysis is executed for steel, aluminium, copper and zinc prices. The split of long and short term relationships via energy distributions and de-noising via wavelet transform are utilized to investigate multiple wavelet coherence. Finally, Vector Autoregressive Moving Average Model (VARMA) is used for forecasting with reduced errors. Conclusion is given at Section 4.

## 2. Methodology

### 2.1. Multiple Wavelet Coherence (MWC)

Multiple Wavelet Coherence(MWC) is an extension from the bivariate to the multivariate case. In this case, the correlation of the variables with each other is taken into account when calculating coherency and phase differences.

The squared multiple wavelet coherency between series $X_1$ and all other series $X_2,\ldots,X_p$ is defined as: [44]

$$R^2_{1(2,3,\ldots,p)} = R^2_{1(q)} = 1 - \frac{M^d}{S_{11} M^d_{11}}. \tag{1}$$

where M is the p x p matrix of all smoothed cross-wavelet spectra $S_{ij}$, which can be shown as:

$$S_{ij} = S\left(W_{x_i x_j}\right) \; (S_{ij} = S^*_{ji}, \; S_{ij} = S(|W_{x_i}|^2)). \tag{2}$$

$$M = \begin{bmatrix} S_{11} & \cdots & S_{1p} \\ \vdots & \ddots & \vdots \\ S_{p1} & \cdots & S_{pp} \end{bmatrix}. \tag{3}$$

$M^d_{ij}$ is the cofactor of the element (i; j) of matrix M and can be represented as:

$$M^d_{ij} = (-1)^{i+j} \det(M^j_i). \tag{4}$$

where $M^j_i$ represents the sub-matrix obtained from M by deleting the i[th] row and the j[th] column and $M^d$ = det M. The complex partial wavelet coherency of $X_1$ and $X_j$ (2 ≤ j ≤ p) is given by;

$$\rho_{1j.q_j} = -\frac{M^d_{j1}}{\sqrt{M^d_{11} M^d_{jj}}}. \tag{5}$$

where $q_j = \{2, \ldots, p\}$. The partial wavelet coherency of $X_1$ and $X_j$ (2 ≤ j ≤ p) is given by;

$$r_{1j.q_j} = \frac{|M^d_{j1}|}{\sqrt{M^d_{11} M^d_{jj}}}. \tag{6}$$

and

$$r_{1j.q_j}^2 = \frac{\left|M_{j1}^d\right|^2}{M_{11}^d M_{jj}^d}. \tag{7}$$

*2.2. Complex Coherence*

Coherence matrix C is composed of smoothed complex wavelet coherencies, while diagonals are all equal to one. In general, it is the p x p matrix of all smoothed complex wavelet coherencies $\rho_{ij}$ ( $\rho_{ij} = \rho_{ji}^*$) and it can be shown as: [43-44]

$$C = \begin{bmatrix} 1 & \rho_{12} \ldots & \ldots \rho_{1p} \\ \rho_{21} & 1 \ldots & \vdots \\ \rho_{p1} & \rho_{p2} \ldots & \ldots 1 \end{bmatrix}. \tag{8}$$

where $\rho_{ij}$ is given as;

$$\rho_{ij} = \frac{S(W_{ij})}{\sqrt{S(|W_i|^2)\,S(|W_i|^2)}}. \tag{9}$$

Then the squared multiple wavelet coherency could be given as;

$$R_{1(q)}^2 = 1 - \frac{C^d}{C_{11}^d}. \tag{10}$$

The complex partial wavelet coherency could be denoted by $\rho_{1j.q_j}$ and can be given as;

$$\rho_{1j.q_j} = -\frac{C_{j1}^d}{\sqrt{C_{11}^d C_{jj}^d}}. \tag{11}$$

While the partial wavelet coherency is the absolute value of the above quantities, the squared partial wavelet coherency is then given as;

$$r_{1j.q_j}^2 = \frac{\left|C_{j1}^d\right|^2}{C_{11}^d C_{jj}^d}. \tag{12}$$

The partial phase difference of $X_1$ over $X_j$ given all other series is calculated with:

$$\emptyset_{1j.q_j} = \text{ArcTan}\left(\frac{I(\rho_{1j.q_j})}{R(\rho_{1j.q_j})}\right). \tag{13}$$

Multiple coherence in terms of partial can be shown as [44]:

$$R_{1(23\ldots p)}^2 = R_{1(q)}^2 = (1 - r_{12}^2)(1 - r_{13.2}^2)\ldots\ldots(1 - r_{1p.23..(p-1)}^2). \tag{14}$$

## 2.3. Four Time Series Case ($X_1$, $X_2$, $X_3$ and $X_4$)

The extension to four variables is executed first in this paper to be able to detect co-movement of those variables. The aim is to be able to detect all possible intra relations between metals to increase the efficiency of forecasting process. In other word, to catch all possible co-movements as much as possible that is effective in metal price, the number of variables is increased up to four.

The first step is to generate the coherence matrix C, which is now 4 x 4 matrix of all smoothed complex wavelet coherencies $\rho_{ij}$:

$$C = \begin{bmatrix} 1 & \rho_{12} & \rho_{13} & \rho_{14} \\ \rho_{21} & 1 & \rho_{23} & \rho_{24} \\ \rho_{31} & \rho_{32} & 1 & \rho_{34} \\ \rho_{41} & \rho_{42} & \rho_{43} & 1 \end{bmatrix}. \quad (15)$$

Since $C_{11}^d$ is needed to calculate the squared multiple wavelet coherency, it is calculated from coherence matrix C as follow:

$$C_{11}^d = \begin{vmatrix} 1 & \rho_{23} & \rho_{24} \\ \rho_{32} & 1 & \rho_{34} \\ \rho_{42} & \rho_{43} & 1 \end{vmatrix}. \quad (16)$$

When we open the above expression in details, we will come up with;

$$C_{11}^d = 1 - \rho_{23}\rho_{32} - \rho_{24}\rho_{42} + \rho_{23}\rho_{34}\rho_{42} + \rho_{24}\rho_{32}\rho_{43} - \rho_{34}\rho_{43}. \quad (17)$$

In equation (17), $\rho_{23}\rho_{32}$ can be expressed as $R_{23}^2$ and in the same logic, $\rho_{24}\rho_{42}$ and $\rho_{34}\rho_{43}$ can be written as $R_{24}^2$ and $R_{34}^2$ respectively. In addition, we can add the remaining two terms to show it as follow;

$$C_{11}^d = 1 - R_{23}^2 - R_{24}^2 - R_{34}^2 + 2R\,(\rho_{23}\rho_{34}\rho_{24}^*). \quad (18)$$

Final step is to calculate $C^d$ from equation (15) to be able to proceed. The detailed derivation of $C^d$ is given in details below:

$$C^d = C_{11}^d - \rho_{12}C_{12}^d + \rho_{13}C_{13}^d - \rho_{14}C_{14}^d. \quad (19)$$

Let us substitute each term in equation (17) in details to find the below detailed formula;

$$\begin{aligned} C^d = {} & 1 - \rho_{12}\rho_{21} - \rho_{13}\rho_{31} - \rho_{24}\rho_{42} - \rho_{14}\rho_{41} - \rho_{34}\rho_{43} + \rho_{12}\rho_{23}\rho_{31} + \rho_{13}\rho_{21}\rho_{32} + \\ & \rho_{12}\rho_{24}\rho_{41} + \rho_{14}\rho_{23}\rho_{32}\rho_{41} - \rho_{13}\rho_{24}\rho_{32}\rho_{41} + \rho_{13}\rho_{34}\rho_{41} - \rho_{12}\rho_{23}\rho_{34}\rho_{41} + \rho_{14}\rho_{21}\rho_{42} - \\ & \rho_{14}\rho_{23}\rho_{31}\rho_{42} + \rho_{13}\rho_{24}\rho_{31}\rho_{42} - \rho_{13}\rho_{22}\rho_{34}\rho_{42} + \rho_{23}\rho_{34}\rho_{42} + \rho_{14}\rho_{31}\rho_{43} - \\ & \rho_{12}\rho_{24}\rho_{31}\rho_{43} - \rho_{14}\rho_{21}\rho_{32}\rho_{43} + \rho_{24}\rho_{32}\rho_{43} + \rho_{12}\rho_{21}\rho_{34}\rho_{43}. \end{aligned} \quad (20)$$

Finally, like we did before in the calculation of $C_{11}^d$, we simplified the formula as follow;

$$C^d = 1 - R_{12}^2 - R_{13}^2 - R_{23}^2 - R_{14}^2 - R_{24}^2 - R_{34}^2 + \rho_{12}\rho_{23}\rho_{31} + \rho_{13}\rho_{21}\rho_{32} +$$
$$\rho_{12}\rho_{24}\rho_{41} + \rho_{14}\rho_{21}\rho_{42} + \rho_{23}\rho_{34}\rho_{42} + \rho_{24}\rho_{32}\rho_{43} + \rho_{14}\rho_{31}\rho_{43} + \rho_{13}\rho_{34}\rho_{41} +$$
$$\rho_{14}\rho_{23}\rho_{32}\rho_{41} - \rho_{13}\rho_{24}\rho_{32}\rho_{41} - \rho_{12}\rho_{23}\rho_{34}\rho_{41} - \rho_{14}\rho_{23}\rho_{31}\rho_{42} + \rho_{13}\rho_{24}\rho_{31}\rho_{42} -$$
$$\rho_{13}\rho_{21}\rho_{34}\rho_{42} - \rho_{12}\rho_{24}\rho_{31}\rho_{43} - \rho_{14}\rho_{21}\rho_{32}\rho_{43} + \rho_{12}\rho_{21}\rho_{34}\rho_{43}. \tag{21}$$

To be able to write the detailed squared multiple wavelet coherency formula, equations (18) and (21) could be placed into below formula which is then used in the numerical calculations.

$$R_{1(234)}^2 = 1 - \frac{C^d}{C_{11}^d}. \tag{22}$$

Equation (22) is used in data analysis part, while calculating squared multiple wavelet coherence for each metal. In the next section, you will find data analysis based on multiple wavelet coherence by using the formulas derived here.

### 3. Data Analysis

The daily prices of steel, aluminium, copper and zinc from May 10, 2010 to May 29, 2014 are analyzed to be able to highlight the interactions via different wavelet analysis. The data contains 1024 observations and the graphical representation is given on Fig.1.

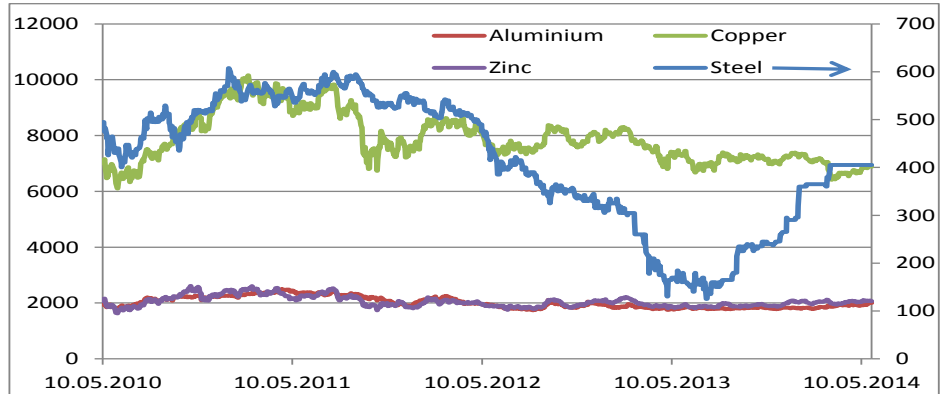

Fig. 1. Steel, Aluminium, Copper & Zinc Prices / Time Series Plot

One of the aims is to split the trend and noise from each other to be able to analyze both short and long term relationships between the mentioned metals. For this purpose, the first step is to apply discrete wavelet transform to each series separately to see the energy distribution between each node. Daubechies wavelet which defines a family of orthogonal wavelets with a wavelet order of three has been used and refinement level is selected as five. In the below table, energy fractions for the first and last two nodes are given to show the energy distribution in details.

Table 1. Discrete Wavelet Transform Energy Fractions

| | | Nodes | | | |
|---|---|---|---|---|---|
| | | {0,0,0,0,0} | {0,0,0,0,1} | {1,1,1,1,0} | {1,1,1,1,1} |
| Precious Metals | Steel | 0.996837 | 0.000344519 | $12.5762*10^{-6}$ | $7.49416*10^{-6}$ |
| | Aluminium | 0.998444 | 0.000218415 | $3.03249*10^{-6}$ | $2.49282*10^{-6}$ |
| | Copper | 0.998366 | 0.000267134 | $3.43233*10^{-6}$ | $3.349*10^{-6}$ |
| | Zinc | 0.997522 | 0.000619824 | $3.8198*10^{-6}$ | $3.47166*10^{-6}$ |

As it can easily be seen from the above table, more than 99% of the energy for each metals is at the first node which is {0,0,0,0,0}. The next step is to apply inverse wavelet transform for each and every metal independently to construct time series for the first and the last node. While applying the inverse wavelet transform for the node {0,0,0,0,0}, the trend of time series is obtained, the last node {1,1,1,1,1} inverse wavelet transform results in the noise part of the time series. The Fig.2 shows the original data series, the data series constructed from the first node (called trend part) and the last node (called noise part) for each metal separately where it can be visualized that the first and the last node represents the trend and noise in the data.

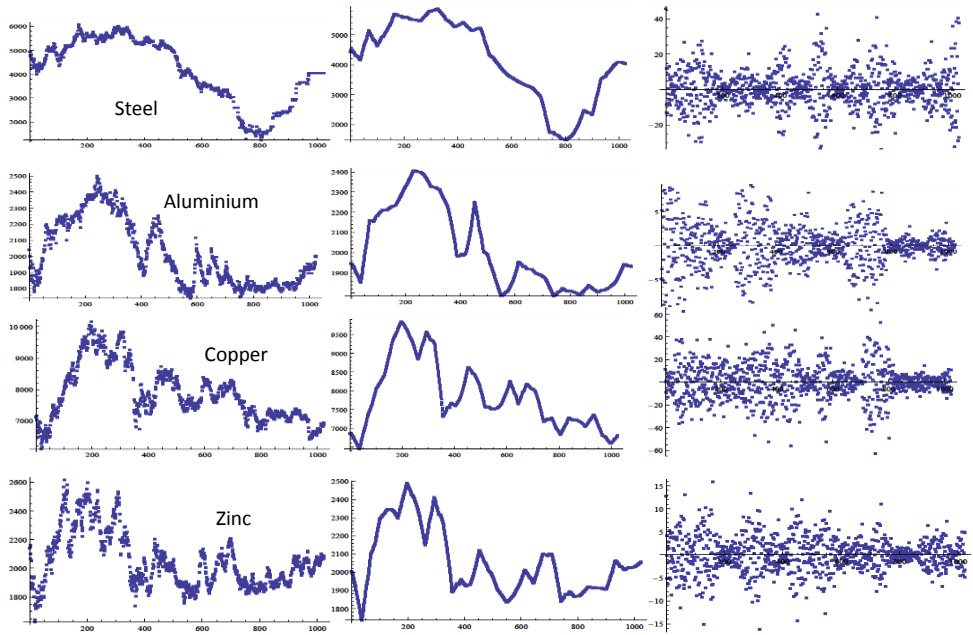

Fig. 2. Original Data, First Node & Last Node [Steel, Aluminium, Copper and Zinc]

*3.1. De-Noising via Wavelet Transform*

Similar to the node approach, the first step make a discrete wavelet transform to each data series where Daubechies wavelet with a wavelet order of three is utilized again. Afterwards, the proper wavelet thresholds are investigated to carry out the best inverse wavelet transform. For this purposes, within all possible methods the one which gives the best Signal to Noise (SNR) and Peak Signal to Noise (PSNR) is selected. Each metal is investigated in terms of both thresholding specification and wavelet indices. Table 2 shows all possible candidates with their SNR and PSNR values.

Table 2. SNR and PSNR Analysis

| \multicolumn{6}{c|}{Steel} | \multicolumn{6}{c}{Aluminium} |
|---|---|---|---|---|---|---|---|---|---|---|---|

| Method | SNR | PSNR | Method | SNR | PSNR | Method | SNR | PSNR | Method | SNR | PSNR |
|---|---|---|---|---|---|---|---|---|---|---|---|
| GCV | 19.0268 | 32.146 | Hard | 29.6812 | 42.8003 | GCV | 16.1682 | 38.0088 | Hard | 23.1076 | 44.9483 |
| GCVLevel | 13.0253 | 26.1445 | Soft | 26.0463 | 39.1655 | GCVLevel | 14.5592 | 36.3998 | Soft | 19.5814 | 41.422 |
| SURE | 76.3848 | 89.504 | Firm | 29.6812 | 42.8003 | SURE | 89.7678 | 111.608 | Firm | 23.1076 | 44.9483 |
| SURELevel | 17.5737 | 30.6929 | PiecewiseGarrote | 28.0496 | 41.1688 | SURELevel | 13.2116 | 35.0523 | PiecewiseGarrote | 21.4045 | 43.2451 |
| SUREShrink | 58.062 | 71.1811 | SmoothGarrote | 31.1464 | 44.2656 | SUREShrink | 64.7428 | 86.5834 | SmoothGarrote | 24.4355 | 46.2761 |
| Universal | 29.6812 | 42.8003 | Hyperbola | 28.9305 | 42.0497 | Universal | 23.1076 | 44.9483 | Hyperbola | 22.28 | 44.1206 |
| UniversalLevel | 9.03471 | 22.1539 | | | | UniversalLevel | 8.22978 | 30.0704 | | | |
| VisuShrink | 26.0463 | 39.1655 | | | | VisuShrink | 19.5814 | 41.422 | | | |
| VisuShrinkLevel | 6.62948 | 19.7486 | | | | VisuShrinkLevel | 6.10688 | 27.9475 | | | |

| \multicolumn{6}{c|}{Copper} | \multicolumn{6}{c}{Zinc} |
|---|---|---|---|---|---|---|---|---|---|---|---|

| Method | SNR | PSNR | Method | SNR | PSNR | Method | SNR | PSNR | Method | SNR | PSNR |
|---|---|---|---|---|---|---|---|---|---|---|---|
| GCV | 15.0744 | 36.1773 | Hard | 20.9312 | 42.0341 | GCV | 26.3685 | 49.0798 | Hard | 18.9852 | 41.6964 |
| GCVLevel | 5.4448 | 26.5477 | Soft | 17.6567 | 38.7597 | GCVLevel | 5.11248 | 27.8237 | Soft | 15.1843 | 37.8955 |
| SURE | 97.6346 | 118.738 | Firm | 20.9312 | 42.0341 | SURE | 96.7007 | 119.412 | Firm | 18.9852 | 41.6964 |
| SURELevel | 18.3739 | 39.4768 | PiecewiseGarrote | 19.5292 | 40.6321 | SURELevel | 27.6758 | 50.387 | PiecewiseGarrote | 17.2388 | 39.9501 |
| SUREShrink | 69.9292 | 91.0321 | SmoothGarrote | 22.3329 | 43.4359 | SUREShrink | 69.2392 | 91.9504 | SmoothGarrote | 20.1505 | 42.8617 |
| Universal | 20.9312 | 42.0341 | Hyperbola | 20.2651 | 41.368 | Universal | 18.9852 | 41.6964 | Hyperbola | 18.1498 | 40.8610 |
| UniversalLevel | 6.40962 | 27.5126 | | | | UniversalLevel | 4.32321 | 27.0344 | | | |
| VisuShrink | 17.6567 | 38.7597 | | | | VisuShrink | 15.1843 | 37.8955 | | | |
| VisuShrinkLevel | 2.48137 | 23.5843 | | | | VisuShrinkLevel | 2.89034 | 25.6016 | | | |

As it can be seen from Table 2, the highest SNR and PSNR values are valid for Smooth Garrote and SURE for each metal. Therefore, the inverse wavelet transform is carried out via those thresholds and the resulted time series are drawn together with original time series in Fig.3.

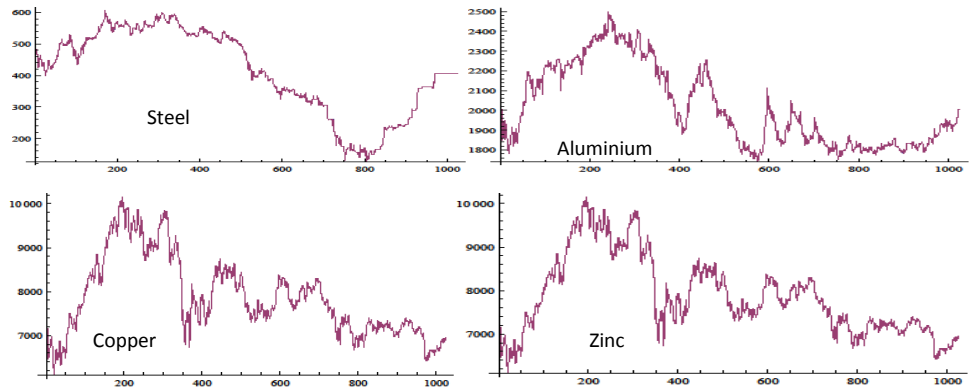

Fig. 3. Original vs De-Noised Time Series

Above figures show us the fact that the thresholds are set in a proper way. Since the time series obtained via both node approach and de-trending, multiple wavelet coherence (MWC) is investigated for four variables in the next section.

### 3.2. Multiple Wavelet Coherence (MWC)

While partial wavelet coherence (PWC) is a technique similar to partial correlation that helps identify the resulting wavelet transform coherence (WTC) between two time series after eliminating the influence of their common dependence, MWC is useful in seeking the resulting WTC of multiple independent variables on a dependent one. In this section, three different multiple wavelet coherence analysis are performed. The first one is related to original time series of steel, aluminium, copper and zinc where each metal is selected as dependent variable and the relationship of dependent one with other independent variables are graphed one by one via the technique explained in Section.2.3. Equation (22) is used to calculate multiple wavelet coherence for each metal. Fig.4 shows the graphical representations of each case.

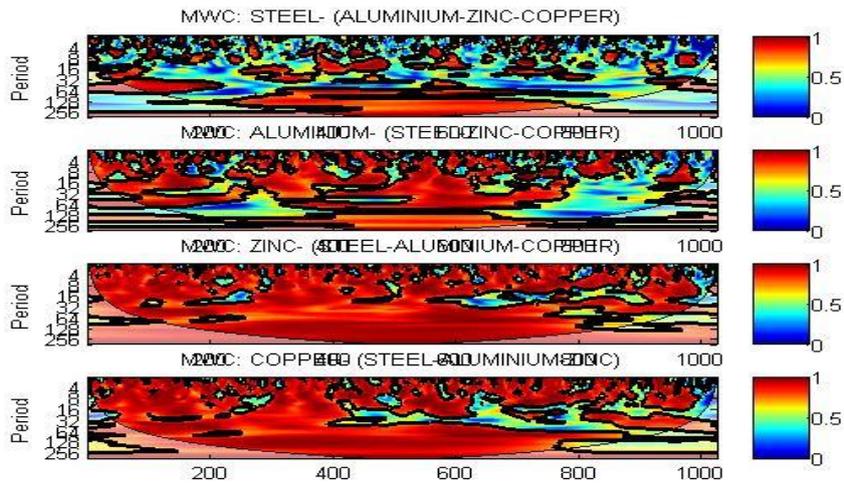

Fig. 4. Original Time Series - Multiple Wavelet Coherence Analysis

The next step is to carry out the same MWC analysis for the time series constructed from the node approach where the trend and noise parts are obtained via wavelet transform methods. While the first chart at Fig.5 shows the relationships between metals trends, the second one describes the noise relationship which is constructed from last node {1,1,1,1,1}.

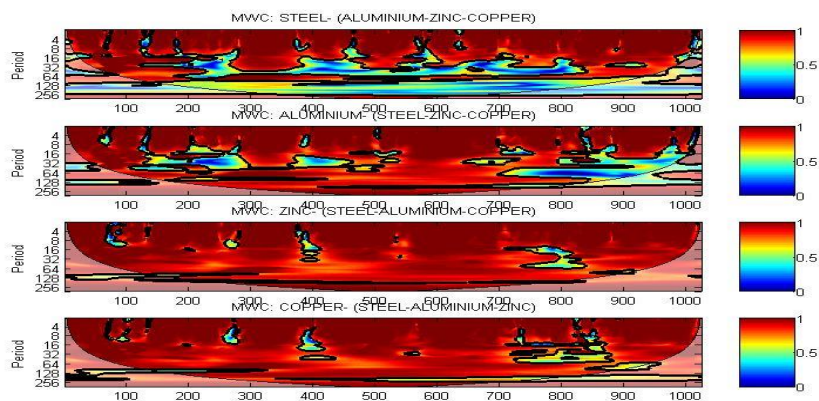

Fig. 5.a. Trend Part of Time Series - Multiple Wavelet Coherence Analysis

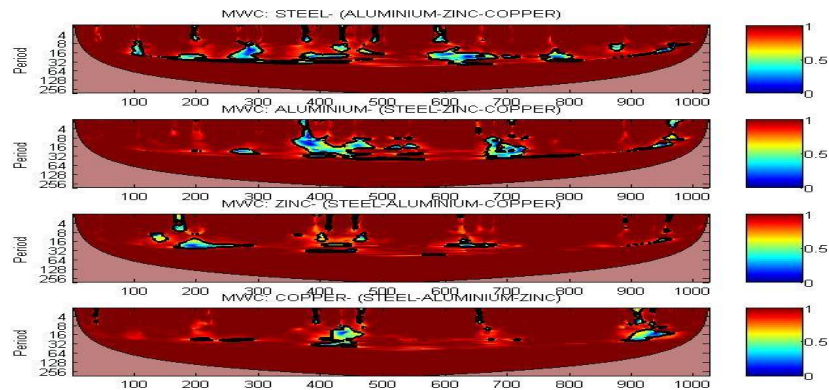

Fig. 5.b. Noise Part of Time Series - Multiple Wavelet Coherence Analysis

Finally, de-noised times series are used in multiple wavelet coherence analysis and the graph shows the relationships between them are given at Fig. 6.

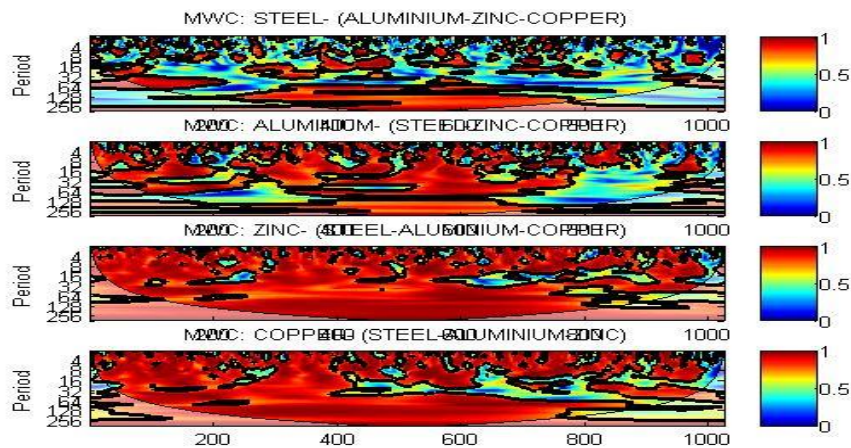

Fig. 6. De-Noised Time Series - Multiple Wavelet Coherence Analysis

The results shows that the short term trend relationships or co-movement of steel price with aluminium, copper and zinc ones are quite correlated; it is not the case for the long term trend. In other words, in a short term period, the steel price moves are quite correlated or linked to price moves of aluminium, copper and zinc. On the other hand, all the others, means aluminium, copper and zinc, are moves quite in line with other three either in short and long term.

Finally, it can easily be stated that correlation between steel, aluminium, copper and zinc prices are quite high between middle range which stands for the end of 2011 and 2012. In the next section, the study will be focused on this time interval and the positive forecasting effect of high correlation between the time series is demonstrated via VARMA model.

## 3.3. Vector Autoregressive Moving Average Model (VARMA)

VARMA models are a powerful tool for producing linear forecasts for a set of time series variables [38]. They utilize the information not only in the past values of particular variable of interest but also allow for information in other related variables. As it is observed at the previous part of the study that the correlation or inter relation between steel, aluminium, copper and zinc prices are quite high between end of 2011 and 2012. Therefore, the related times series between 14.11.2011 and 16.11.2012 are studied via autoregressive moving average models (both ARMA and VARMA). The steel, aluminium, prices and zinc are multiplied by 10, 2 and 2 respectively; while copper prices are divided by 2 to be able both to see them in the same axis and work on numerically similar scale figures. The data contains 256 observations and the graphical representation is given on Fig.7.

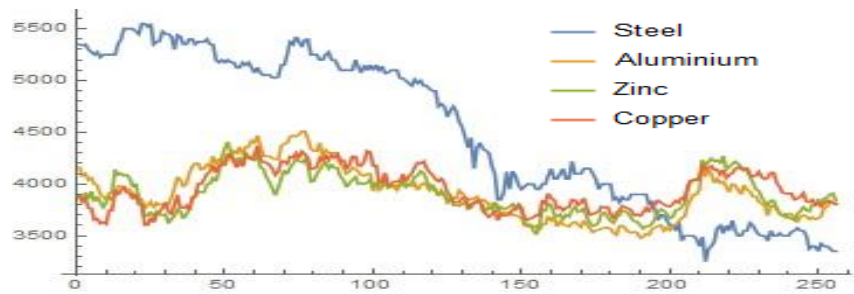

Fig. 7. Steel, Aluminium, Copper & Zinc Prices / Time Series Plot

While ARMA(1,1) model is applied to each metal separately, VARMA(1,1) model is studied for those four metal. The output of each model is utilized to predict future thirty prices. On Fig.8, all results together with original series and predicted ones are presented.

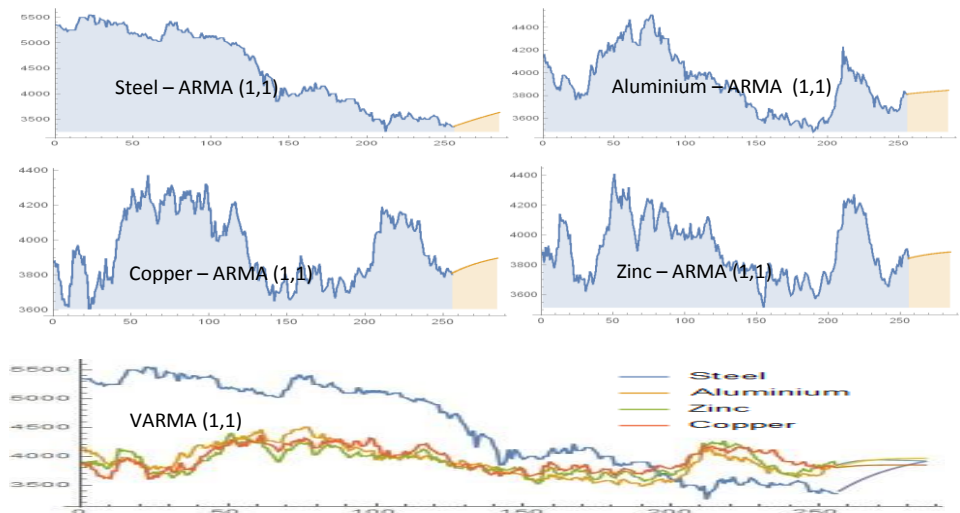

Fig. 8. ARMA (1,1) & VARMA(1,1) Model Outputs

The next step is to check that whether VARMA(1,1) is better than ARMA(1,1) in terms of MSE. On Fig.9, the graphs show the mean square error bands with 5% confidence level for each metal. It can be easily stated that vector approach yields better estimates with respect to univariate ones, because of the supportive benefit of correlation between the variables.

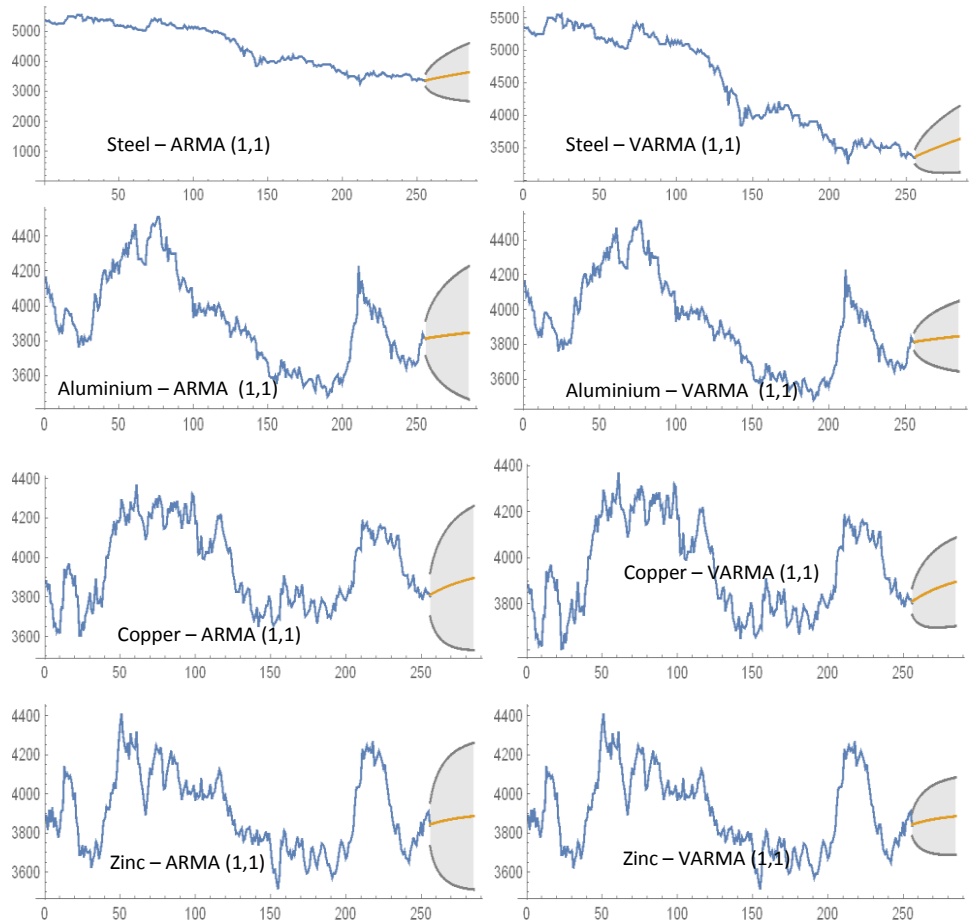

Fig. 9. Mean Square Error - ARMA (1,1) versus VARMA(1,1) 5% Confidence Levels

## 4. Conclusion

In this study, the daily prices of steel, aluminium, copper and zinc between 10.05.2010 and 29.05.2014 are analyzed via different wavelet analysis approaches. Node and de-noise approaches are utilized to be able to understand the dynamics of price moves for mentioned metals in the time-frequency space. The aim is to identify the co-movements together with long and short term trend relationships. The results shows that in the short term, there is a high relationship between those metals; only for steel, it is not valid in the long term.

The next step is to check the positive effect of those co-movements in the estimation process. For this purpose, the data band where the correlations between steel, aluminium, copper and zinc prices are quite high is selected to further examine via ARMA and VARMA models. Metal prices between 14.11.2011 and 16.11.2012 are selected as time series for those analysis and thirty new prices are estimated with both ARMA(1,1) and VARMA(1,1) models. The precision of estimates are compared by using mean square errors with 5% confidence level and the results clearly showed that vector analysis contributes extra on top. In other word, the prediction power could be easily increased and the prediction errors could be reduced by using autocorrelations, partial autocorrelations and cross correlations between the variables for model specification.

It is clearly shown that combining MWC approach in forecasting process makes the estimation outputs better and improve the precision of forecasts. In other word, dynamic co-movement detection via wavelet coherency analysis in the determination of VARMA modelling time intervals enables to decrease forecasting errors appreciably. The results of the data analysis confirmed the benefit of co-movement analysis using wavelet coherence method. Results also showed the validity of four variables WTC analysis and the improvement of ARMA's forecasting power via checking for VARMA and forecasting through it. In our opinion, it is a preliminary step in metal price modelling and forecasting. The method could be utilized efficiently in the studies and researches focused on metal sector.